# Possible Application of Wavefront Coding to the LSST[1]

Willy Langeveld

*Stanford Linear Accelerator Center*
*Stanford University*
*Stanford, CA*

## ABSTRACT

Wavefront Coding has been applied as a means to increase the effective depth of focus of optical systems. In this note I discuss the potential for this technique to increase the depth of focus of the LSST and the resulting advantages for the construction and operation of the facility, as well as possible drawbacks. It may be possible to apply Wavefront Coding without changing the current LSST design, in which case Wavefront Coding might merit further study as a risk mitigation strategy.

---

[1] Work supported in part by the Department of Energy contract DE-AC03-76SF00515

# 1. Introduction

Wavefront Coding is a technique, invented by W. Thomas Cathey and Edward R. Dowski, that combines modified optics with digital signal processing (DSP) to produce an after-DSP image of improved quality compared to unprocessed images obtained using conventional optics[1]. In particular, such systems can be used to extend the depth of field and depth of focus of optical imaging systems[2]. Certain applications to space-based astronomical telescopes have been discussed in the literature[3]. A discussion of the recent trend to integrate optics and electronics can be found in reference 4.

The LSST is a proposed ground-based optical telescope with primary mirror which has a diameter of 8.4 m and a system f-number of 1.25[5]. Since the proposed pixel size of the camera is 10 microns, the depth of focus of the telescope is roughly 25 microns. Because of this, the tolerances on the alignment and positioning of the various telescope components are very tight. As an example, the tolerances for the vertical positioning of the ~200 image sensors on the focal plane array (FPA) are on the order of 10 microns peak to valley. To achieve this precision, a large and expensive effort is being mounted to ensure that materials and fixtures will have the required properties. In addition, various methods are under study to measure the FPA flatness in the lab, and monitor it in situ. Operationally, alignment and focusing will also require great precision.

Each tolerance contributes to the overall project risk. The tighter the tolerance, the more it will contribute to the risk. Since the tightness of many tolerances in the LSST design are largely due to the small depth of focus, it seems worth while studying the implications of Wavefront Coding to increase the depth of focus and thereby mitigating the risk. Note, that I do not advocate that any particular tolerance could, or should, be relaxed. Rather, the argument is that in case some tolerances are not met in the final product, there may be a way to recover the design imaging quality through this method. Tolerances can reasonably only be relaxed as part of a complete redesign of the telescope that incorporates Wavefront Coding as an integral part of the design, a task we do not contemplate here.

In this note I give a short overview of Wavefront Coding and a discussion of how it might be applied to the LSST and the potential problems associated with the technique.

# 2. Wavefront Coding

Wavefront Coding is accomplished by putting a phase mask at a pupil in an imaging system. In practice the effect of the phase mask is realized on the surface of an existing optical element. The phase mask has to be chosen such that its effects on the raw image can be unfolded from the raw image data without ambiguities. Phase masks modify the point spread function (PSF) in certain ways depending on the details of the mask. It is possible to design phase masks for specific purposes. As an example, a cubic phase mask will cause the PSF of a perfectly focused point source to closely resemble the PSF of point source that is a certain amount out of focus. As it turns out, digital signal processing of the raw image will reconstruct, within certain limits, both objects as perfectly focused in the processed image.

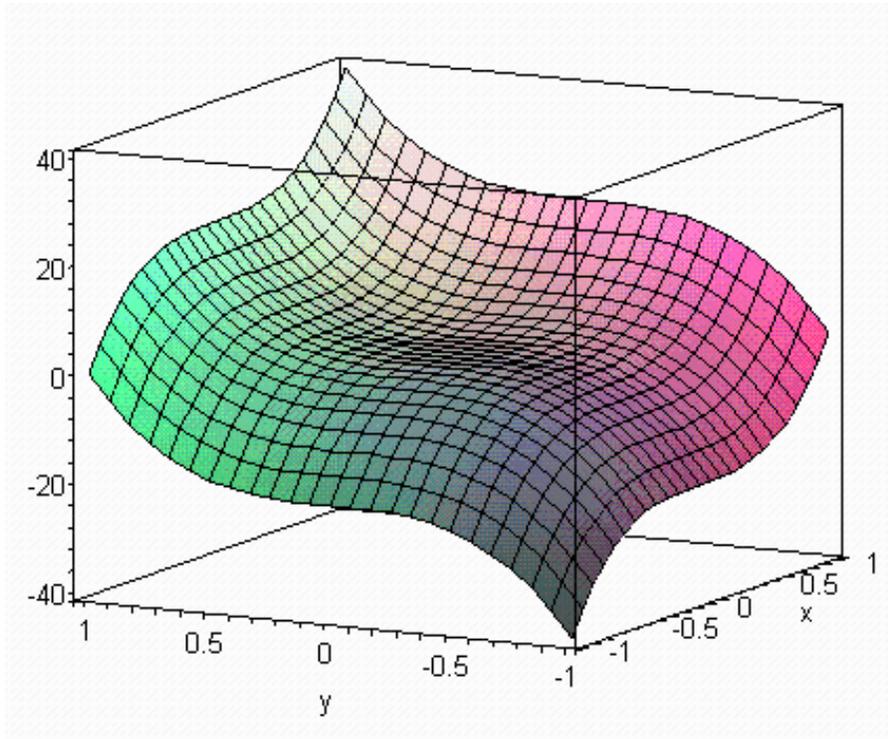

**Figure 1. A simple cubic phase mask.**

The simplest example of a cubic phase mask is given by the equation:

$$z = \alpha (x^3 + y^3)$$

Where x and y are in normalized coordinates ($-1 \leq x, y \leq 1$). A picture of such a mask (with $\alpha = 20$) is shown in Figure 1. This mask happens to be separable in x and y; in general that is not a necessary condition. Separable masks do have the advantage that the deconvolution can be performed in two steps, one processing the rows and the other processing the columns, but only for rectangular apertures. In general, deconvolution is more compute-intensive.

The PSFs associated with this type of mask are shown in Figure 2, adapted from reference 1. Figure 2A shows the PSFs of a point source using a conventional imaging system in focus (left) and out of focus (right.). Figure 2B shows the PSFs for the same two point sources using Wavefront Coding with a cubic phase mask. Figure 2C shows the PSFs of the wavefront coded images after deconvolution. Clearly, a point source that was severely out of focus using conventional optics is perfectly in focus using Wavefront Coding.

In practice one hardly ever uses a simple cubic phase mask such as the one above. For most applications there are much better phase functions. Typical phase functions add between 2 and 10 wavelengths worth of phase to the wavefront[6].

A number of other examples can be found on the website[7] of CDM Optics, a company that specializes in wavefront coded optics consulting, simulation and licensing. It will not come as a surprise that the principal authors of references 1 and 2 are founders of the company.

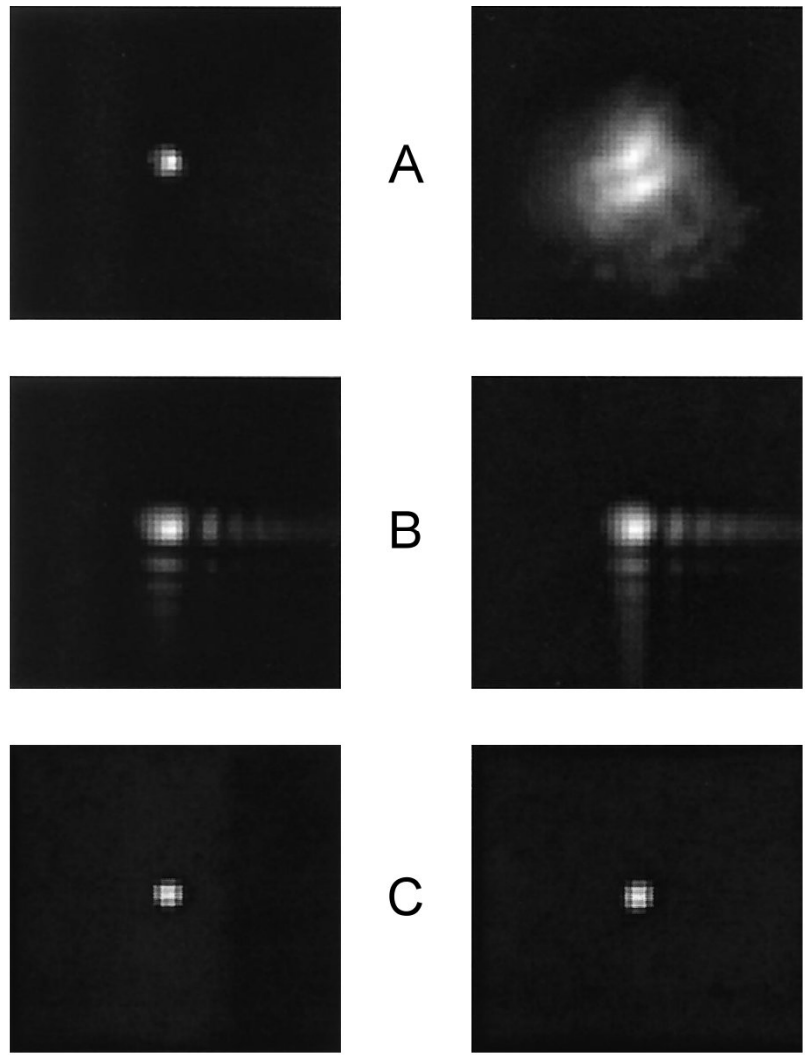

Figure 2. PSFs of in-focus (left) and out-of -focus (right) point sources: (A) using conventional optics, (B) using wave-front coded optics, raw image and (C) using wave-front coded optics, processed image. Images taken from reference 1.

## 3. Possible Application to the LSST

Wavefront Coding requires the presence of a phase mask at a pupil in the imaging system. Usually, the imaging system is designed such that it has a pupil somewhere with a manageable diameter in order to limit the difficulty in fabricating the phase mask.

The current design of the LSST has only one true pupil, the entrance pupil, i.e. the primary mirror. While it might be possible to polish a phase mask into the primary, I would consider that a change of the LSST design, a possibility I will not consider here since this note only considers Wavefront Coding as a risk mitigation strategy of the current design.

On the other hand, the primary mirror is an actuated mirror, and one might consider the possibility of deliberately distorting the mirror using the actuators so as to produce a suitable phase mask. Alternatively, the secondary mirror is close to a pupil and also actuated, so perhaps it too could be actuated in such a way as to include a phase mask. The details are really a matter that can only be settled by a detailed simulation of the entire telescope with the addition of a phase mask. This falls outside the scope of this note, however.

## 4. Benefits of Wavefront Coding

Assuming one can use the active optics of the primary and/or secondary mirrors to provide a phase mask, there are several potential benefits to the LSST. Tolerances on focal plane flatness and overall telescope alignment would be much less strict. If Wavefront Coding were to be shown as feasible and accepted as a risk mitigation strategy, one could decide two currently unresolved issues in the camera design: the "rafts" holding the sensors could be mounted directly onto the integrating structure in the camera without vertical positioning devices and there would be no need for any in-situ flatness verification systems.

Wavefront Coding also improves system performance with regard to atmospheric turbulence. All Zernike terms except for the ones describing a tilt would improve.

Obviously, focusing would be easier.

A smaller benefit would be that since point sources are spread out across more pixels by the phase mask, blooming of foreground stars would be slightly improved.

One could still use wavefront sensors for alignment purposes: one would just have to subtract out the effects of the phase mask. In fact, one would probably use the wavefront sensors to guide the set up of the phase mask in M1 and/or M2.

Unfortunately, it is difficult to estimate how much the improvement in the depth of focus would be without doing a detailed simulation. The phase mask must be chosen carefully to provide the desired benefits without inducing distortions or causing excessive signal-to-noise losses. There are families of phase masks that do not introduce distortions, but some do not have a constant gain in spatial frequency, and may lead to some correlated noise[6].

Finally, it should be noted that a wavefront coded system is even somewhat tolerant to errors in the phase mask itself.

## 5. Drawbacks of Wavefront Coding

The primary drawback of Wavefront Coding is a small reduction in signal-to-noise ratio at the traditional best focus. Of course, this assumes that the telescope is perfectly aligned and in perfect focus. For many situations it is quite possible that the signal-to-noise ratio of a wavefront coded system is higher than that of a somewhat misaligned conventional setup.

Wavefront Coding also requires a two-dimensional deconvolution step. It is probably possible to do this directly in hardware as part of the signal processing chain. Systems have been designed for processing the images of a 20k x 20k, 12-bit imaging system using four FPGAs[6], and larger systems such as the LSST (60k x 60k, 16-bit) could probably be done in a similar way.

Another drawback of Wavefront Coding is that there are some possible edge effects. Since the PSFs extend across more pixels, near the edges of the sensors there will be a slight loss of information due to the gaps between them. In the bulk of the sensor array, software can pick up the additional information from the neighboring sensors during the stitching process, so the effect there is not very large. At the outer edges of the focal plane, partial reconstruction could still be performed, but it is more likely that one would just reduce the fiducial area of the FPA by a very small amount.

## 6. Summary

In this note I have described a potentially interesting technique, Wavefront Coding, which could be used as a risk mitigation strategy for the LSST. A large number of dimensions have very tight tolerances, many of which are significant risk factors for the project. Wavefront Coding could be used if it turns out that it is impossible to meet one or more of the tolerances. It may be possible to use the Wavefront Coding "fix" without modifying the current design of the LSST, by using the active optics of the primary and/or secondary mirrors to add a suitable phase mask.

I wish to credit Dick Blankenbecler for suggesting the potential relevance of Wavefront Coding to the LSST and I thank him for helpful discussions. I also thank Tom Cathey and Ken Kubala for answering many of my questions.

---

[1] W. Thomas Cathey, Edward R. Dowski, "New Paradigm for Imaging Systems", Applied Optics **41**, p. 6080 (2002).
[2] Edward R. Dowski, W. Thomas Cathey, "Extended Depth of Field through Wave-Front Coding", Applied Optics **34**, p. 1859 (1995).
[3] Kenneth Kubala, Edward Dowski, James Kobus, Bob Brown, "Design and Optimization of Aberration and Error Invariant Space Telescope Systems", Proceedings of the SPIE **5524**, pp. 54-65 (2004).
[4] Joseph N. Mait, Ravi Athale, Joseph van der Gracht, "Evolutionary Paths in Imaging and Recent Trends", Optics Express **11**, p. 2093 (2003).
[5] W. Althouse et al., "The Large Synoptic Survey Telescope (LSST)", Letter of Intent to the EPAC at SLAC. See also the "Proposal for LSST R&D", a SLAC Report currently being prepared.
[6] Ken Kubala, CDM Optics, *private communication*.
[7] http://www.cdm-optics.com